\documentclass[aps,prb,amsmath,amsfonts,amssymb,showpacs,superscriptaddress,floatfix,twocolumn]{revtex4-1}
\usepackage{times}
\usepackage{epsfig}
\usepackage{hyperref}
\usepackage{bm}
\newcommand{\ii}{\mathrm{i}}
\newcommand{\dd}{d}

\newcommand{\cS}{c_\text{S}}
\newcommand{\cL}{c_\text{L}}

\newcommand{\pv}{\mathop{\text{p.v.}}}
\newcommand{\Pf}{\mathop{\text{Pf}}}
\renewcommand{\Re}{\mathop{\text{Re}}}

\begin{document}
\title{Reply to ``Comment on `Dynamic Peierls-Nabarro equations for elastically isotropic crystals'\,''}
\author{Yves-Patrick \surname{Pellegrini}}
\email{yves-patrick.pellegrini@cea.fr} \affiliation{CEA, DAM, DIF, F-91297 Arpajon, France.}
\date{1 December 2010}
\begin{abstract}
The Comment by Markenscoff that criticizes a recent dynamic extension of the Peierls-Nabarro equation [Y.-P.~Pellegrini, Phys.\ Rev.\ B \textbf{81}, 024101 (2010)] is refuted by means of simple examples that illustrate the interest of using an approach based on generalized functions to compute dynamic stress fields.
\end{abstract}
\maketitle
\section{Introduction}
A Comment by Markenscoff\cite{MARK10} [hereafter referred to as (M)] criticizes several aspects of my paper\cite{PELL10a}  [referred to as (P)] on the de\-ri\-va\-tion of dynamic extensions to the static Peierls-Nabarro equations. One remark\cite{MARK10} concerns my account of the author's work in the concluding Section of (P). Admittedly, the expression ``extremely singular'' I employ there is improper, for it might be interpreted as a statement that the problem involves strong \emph{non-regularizable} singularities, which it does not. To address the remaining points, I refer to Eshelby's seminal paper\cite{ESHE53} on dynamic dislocation motion by (E), and use equation numbers preceded by `P', `M' of `E' to refer to equations in (P), (M), or (E). Unless otherwise stated, other equation numbers relate to equations in the present Reply.

\section{Discussion}
\subsection{Eshelby's dynamic PN equation}
\label{sec:pn}
Ref.\ (P) focuses on obtaining dynamic equations of the Peierls-Nabarro (PN) type (to be solved numerically) to determine the time-varying slip $\eta(x,t)$ on the slip plane $y=0$ of a planar dislocation with extended core, subjected to an arbitrary time-dependent loading. The periodic pull-back force is arbitrary as well. Accordingly, no assumption on the slip function must be made except smoothness. In this context I claim in (P) is that Eshelby's dynamic PN equation for screws (E21) misses the term (proportional to) $-\partial\eta/\partial t$ featured by (P35). This is seen most directly from Appendix C1 of (P), in which what I call (after Rosakis\cite{ROSA01}) ``Weertman's equation'', namely, the constant-velocity limit (P45) of the dynamic PN equation in the co-moving frame, is easily retrieved from (P35). This cannot be done with (E21), because the term $-\partial\eta/\partial t$ proves essential in the process.\cite{PELL10a} Whereas the kernel $K(x,t)$ represents the part of the cylindrical stress waves generated by the dislocation, that sweeps the slip plane, $-\partial\eta/\partial t$ accounts for outward emission of updating waves collected at their incipience. On the basis of its negative sign\cite{ROSA01} it can be interpreted
(superficially, at least) as a radiative loss term. However, as shown below in Section \ref{sec:dirac}, its operative role is to prevent a spurious stress term to show up at the dislocation core.

Although claiming in (M) that (E21) is correct, Markenscoff offers no convincing supporting argument. The self-force used by Al'shitz et al., namely (E28), is derived by Eshelby from (E26) as part of the equation of motion (EoM) that governs the time evolution of the dislocation position. This EoM assumes an arctangent slip function. Eshelby does \emph{not} derive (E26) from (E21). Since (P) does not address EoM issues (dealt with elsewhere\cite{PILL07,PELL10b}) but a more general question, the paper by Al'shitz et al.\ put forward in (M) is irrelevant here. It should be noted, however, that (E26) \emph{can} be obtained from (P35).\cite{PELL10b} Moreover in retrieving,\cite{MARK01} by assuming an arctangent slip function, the specific \emph{solution} to Weertman's equation that corresponds to a sine pull-back force ---rather than the latter general equation \textit{per se}, contrary to what (M) states--- Markenscoff appeals to non-zero $y$ values.\cite{MARK01} This is not allowed in the context of (E21), so that this calculation is inconclusive with regard to the latter equation.

\subsection{On single-valuedness and distributions}
Prior to (P), DeWit\cite{DEWI73} emphasized that the multivalued character of static displacements associated to individual dislocations can be disambiguated by adding a distribution\cite{SCHW66} (or ``generalized function''\cite{GELF64}) part located on a Volterra cut. Whereas using a single- or a multi-valued displacement field can be seen in statics as a matter of choice,\cite{NOTE2} made possible by an underlying gauge-invariant structure,\cite{KLEI08} a single-valued treatment such as that used in (P) is perhaps more natural in dynamics. Indeed, it is unclear how to extend to dynamics the physical arguments that have been put forward in statics to justify, in connection with mutivaluedness, the arbitrariness of this discontinuity surface.\cite{KLEI08} Also, the surface spanned by a moving individual dislocation line, identified in (P) with the cut, becomes in dynamics a physical observable related to the knowledge of past trajectory. The latter is inherent to dynamics since past motion is used to compute retarded fields.

Obviously, the necessary term $-\partial\eta/\partial t$ is linked to this choice of jumping plane. The question of whether the cut might, after all, be arbitrary in a dynamic theory of dislocation \emph{lines} should be examined carefully, perhaps in the light of gauge-related considerations.\cite{KLEI08,LAZA10} Fig.\ 1 in (M) suggests such an arbitrariness, but it is noted that the result it relates to concerns the \emph{short-time} response of a \emph{Volterra} dislocation. At any rate, the choice made in (P) is appropriate\cite{NABA67} to extended planar
(i.e., Somigliana) dislocations such as in the PN equation.

Inasmuch as one adopts Mura's eigenstrains approach\cite{MURA87} as in (P) the cut constitutes the support of the plastic eigenstrains, which must be properly accounted for when computing fields on the slip plane.\cite{KLEI89} Distribution theory can then be appealed to in order to make calculations without the need for a limiting process across the surface of discontinuity.\cite{ESTR85} In (P) in particular, Fourier transforms (FT) are always implicitly interpreted as generalized functions.

\subsection{Inclusion vs.\ dislocation formulations}
\begin{figure}
\includegraphics[width=5cm]{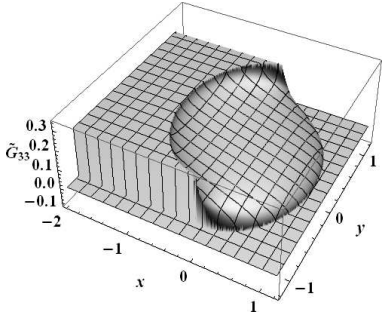}
\caption{\label{fig:fig1} Function $\widetilde{G}_{33}(x,y,t)$ at $t=1$.}
\end{figure}

Two writings of the displacement $\mathbf{u}$ generated by an ei\-gen\-dis\-tor\-tion $\beta^*$, representing an inclusion embedded in a host medium of infinite extent, are possible. To follow (M), we call them the ``inclusion'' and ``dislocation''  representations. In a plane problem these are, with $\Delta{\mathbf{x}}$ $=$ $\mathbf{x}-\mathbf{x}'$ and $\Delta t=t-t'$,
\begin{eqnarray}
&&
\label{eq:uincl}
u_i^{inc}(\mathbf{x},t)=-\!\!\int\!\!\dd t'\,\dd^2\!x'G_{ij}(\Delta{\mathbf{x}},\Delta t)C_{jklm} \beta^*_{lm,k}(\mathbf{x}',t'),\\
\label{eq:udisl}
&&u_i^{dis}(\mathbf{x},t)=-\!\!\int\!\!\dd t'\,\dd^2\!x'G_{ij,k}(\Delta{\mathbf{x}},\Delta t)C_{jklm}
\beta^*_{lm}(\mathbf{x}',t'),
\end{eqnarray}
in which $G_{ij}$ is the Green's function, $C_{ijkl}$ is the tensor of elastic moduli, and where integration over $\mathbf{x}'$ is on infinite space. Their eventual reduction to ``surface'' integrals depends on the precise form of $\beta^*$. As far as $\mathbf{x}'$ is concerned and if $\beta^*$ is bounded by some constant,  $t$ and $t'$ stand as mere parameters, and boundary contributions at infinity can be ignored. Equation (M8) is obtained from (\ref{eq:udisl}) by Mura\cite{MURA87} (see p.\ 351). Expressions (\ref{eq:uincl}) and (\ref{eq:udisl}) differ only by a partial integration with respect to $\mathbf{x}'$, which makes them  formally equivalent. A FT makes this obvious:\cite{MURA87}
\begin{equation}
\label{eq:ufour}
u_i(\mathbf{k},t)=-\ii\int\dd t'\!\!\int \frac{\dd^2\!k}{(2\pi)^2}k_k G_{ij}(\mathbf{k},t-t')C_{jklm}\beta^*_{lm}(\mathbf{k},t'),
\end{equation}
and the factor $\ii k_k$ can be interpreted as a derivative pertaining either to $G_{ij}$ or to $\beta^*_{lm}$. Calculations have been done in (P) using (\ref{eq:ufour}), but it is instructive to compute $u_i$ in direct space with (\ref{eq:uincl}) and  (\ref{eq:udisl}). All formulations are found \emph{equivalent in the transient regime} contrary to what the first paragraph of (M) suggests. To show this, consider a time-varying eigendistortion localized on a surface S of normal $\mathbf{n}(\mathbf{x})$ by means of a surface distribution $\delta_S(\mathbf{x})$, suitable to represent dislocations.\cite{MURA87} Specifically,
$\beta^*_{ij}(\mathbf{x},t)=b_i(\mathbf{x},t) n_j(\mathbf{x})\delta_S(\mathbf{x})$, in which $b_i$ is the local Burgers vector. Then, in the ``inclusion'' formulation the source term is the sum of one single- and one double-layer:
\begin{equation}
\partial_k\beta^*_{ij}=\partial_k (b_i n_j)\delta_S(\mathbf{x})+b_i n_j\,\partial_k \delta_S(\mathbf{x}).
\end{equation}
For a Volterra screw of slip plane $y=0$ one has $\beta^*_{ij}(\mathbf{x})=b(t)\delta_{i3}\delta_{j2}\theta(-x)\delta(y)$ in which $\theta$ is Heaviside's function. The part of $\delta_S(\mathbf{x})$ is played by $\delta(y)$, and that of $\mathbf{b}$ by $b_i(\mathbf{x},t)=b(t)\theta(-x)\delta_{i3}$. Thus
\begin{equation}
\partial_k\beta^*_{ij}(\mathbf{x},t)=b(t)\delta_{i3}\delta_{j2}[-\delta(x)\delta(y)\delta_{k1}
+\theta(-x)\delta'(y)\delta_{k2}].
\end{equation}
For a screw in isotropic elasticity the single-layer contribution vanishes ($\mu$ is the shear modulus):
$C_{jklm}\partial_k\beta^*_{lm}(\mathbf{x},t)=\mu b(t)\theta(-x)\delta'(y)\delta_{j3}$. Focusing on the elementary time-\-de\-pen\-dent solution (P28), which lies at the root of the derivation of the term $-\partial\eta/\partial t$ in (P35), requires taking $b(t)=b\delta(t)$.  The dis\-tri\-bu\-tion\-al derivative in $\delta'(y)$ then entails
\begin{eqnarray}
&&u_3^{inc}(\mathbf{x},t)=-\mu b\int \dd t'\delta(t')\int_{-\infty}^0\!\!\!\!\dd x'\int\dd y'\nonumber\\ &&\hspace{3cm}G_{33}(x-x',y-y',t-t')\delta'(y')\nonumber\\
\label{eq:uincl2}
&&=-\mu b\partial_y \int_{-\infty}^0\dd x'\,G_{33}(x-x',y,t).
\end{eqnarray}
On the other hand, the ``dislocation'' formalism directly yields
\begin{eqnarray}
\label{eq:udisl2}
u_3^{dis}(\mathbf{x},t)=-\mu  b\int_{-\infty}^0\dd x'\, \partial_y G_{33}(x-x',y,t).
\end{eqnarray}
Whereas in the ``dislocation'' formulation, by definition, the derivative $\partial_y$ takes place inside the integral over $x'$, an ambiguity might subsist in the ``inclusion'' formulation, because whether the result should involve $\partial_y\int \dd x'\ldots$, or $\int \dd x' \partial_y \ldots$, or either indifferently, depends on whether the integrals over $x'$ and $y'$ in (\ref{eq:uincl2}) can be harmlessly interchanged or not. In writing the final expression of (\ref{eq:uincl2}) it has been arbitrarily assumed that the integral over $x'$ was done first. Simplifying the writing of the integrals in the right-hand sides (rhs) of (\ref{eq:uincl2}) and (\ref{eq:udisl2}) by an obvious change of variables in $x'$, and introducing
\begin{equation}
\widetilde{G}_{33}(x,y,t)\equiv\int_{x}^{+\infty} \dd x'\,G_{33}(x',y,t),
\end{equation}
the proof of the equivalence consists in showing that
\begin{equation}
\partial_y \widetilde{G}_{33}(x,y,t)=\int_{x}^{+\infty}\dd x'\, \partial_y G_{33}(x',y,t),
\end{equation}
where, with $r=\sqrt{x^2+y^2}$, the Green's function reads\cite{BART89} ($\cS$ is the shear wave velocity)
\begin{equation}
G_{33}(x,y,t)=\frac{\theta(t)\cS}{2\pi\mu}(\cS^2 t^2-r^2)^{-1/2}_+.
\end{equation}
In this writing, the generalized function $x^\alpha_+=x^\alpha$ if $x>0$, and $0$ otherwise, has been introduced to denote what can be written $\theta(\cS^2 t^2-r^2)/\sqrt{\cS^2 t^2-r^2}$.
Then, for $\cS t\not =|y|$,
\begin{eqnarray}
&&\widetilde{G}_{33}(x,y,t)=\frac{\theta(t)\cS}{2\pi\mu}\int_{x}^{+\infty} \dd x'\,(\cS^2 t^2-x^{\prime 2}-y^2)^{-1/2}_+\nonumber\\
&=&\frac{\theta(t)\cS}{2\pi\mu}\theta(\cS t -|y|)\int_{\frac{x}{\sqrt{\cS^2 t^2-y^2}}}^1 \dd u \frac{\theta(1-|u|)}{\sqrt{1-u^2}}\\
&=&\frac{\theta(t)\cS}{2\pi\mu}\left[\pi\theta_D(x,y,t)+\theta(\cS t-r)\arccos\frac{x}{\sqrt{\cS^2 t^2-y^2}}\right],\nonumber
\end{eqnarray}
where $\theta_D(x,y,t)$ stands for the characteristic function of the spatio-temporal domain $D$ delimited by the simultaneous constraints $\{x<0,r>\cS t,|y|<\cS t \}$, equal to $1$ in $D$ and to $0$ elsewhere. The function $\widetilde{G}_{33}$ is represented in Fig.\ \ref{fig:fig1} at $t=1$ (with $\mu=1$, $\cS=1$). The plateau is supported by domain $D$. On the circle $x^2+y^2=\cS^2 t^2$, $\widetilde{G}_{33}(x,y,t)$ is equal to $\theta(-x)/4$ and connects continuously to the plateau on the left semi-circle ($x<0$). Discontinuities are present along the $Ox$ axis at $x=0$ for $y=\pm \cS t$, and along the $Oy$ axis at $y=\pm \cS t$ for $x<0$. The latter are responsible for the Dirac term $\delta(\cS t-|y|)$ in (P28), this equation being retrieved from taking the $y$-derivative of the above expression of $\widetilde{G}_{33}$. The above shows that the Dirac sheets at $y=\pm\cS t$ arise for $x<0$ as the envelope of circular waves emitted at each point of the perturbed region $\{x<0,y=0\}$.

Turning now to the ``dislocation'' formulation, one has\cite{GELF64} $(x^\alpha_+)'=\alpha x_+^{\alpha-1}$  ($-1<\alpha<0$) in the sense of generalized functions, or more formally\cite{SCHW66} $\alpha\Pf x_+^{\alpha-1}$ with Pf denoting Hadamard's finite part, in the sense of distributions (which requires using test functions). Then, for $\cS t\not =|y|$,
\begin{eqnarray}
&&\int_{x}^{+\infty} \dd x'\,\partial_y G_{33}(x',y,t)\nonumber\\
\label{eq:fp32}
&&=\frac{\theta(t)\cS}{2\pi\mu}y\int_{x}^{+\infty} \dd x'\,(\cS^2 t^2-x^{\prime 2}-y^2)^{-3/2}_+.
\end{eqnarray}
The integral in (\ref{eq:fp32}) can be computed conveniently without appealing to test functions, by using a representation of the generalized function similar to those found in Ref.\ \onlinecite{GELF64}
\begin{equation}
(\cS^2 t^2-x^2-y^2)^{-3/2}_+=\lim_{\epsilon\to 0}\Re[\cS^2 t^2-(x+\ii\epsilon)^2-y^2]^{-3/2}.
\end{equation}
In the limit, the real part operator $\Re$ sets the rhs.\ to zero for $\cS^2 t^2-x^2-y^2<0$, as must be. Carrying out the integral yields, for any nonzero real constant $a$,
\begin{eqnarray}
&&I\equiv\int_{x}^{+\infty}\hspace{-1ex} \frac{\dd x'}{[a-(x'+\ii\epsilon)^2]^{3/2}}=-\frac{1}{a}\left[\frac{x+\ii\epsilon}{\sqrt{a-(x+\ii\epsilon)^2}}
-\ii\right]\nonumber\\
\label{eq:epsfinite}
&&{}+\theta(-x)\theta(-a-\epsilon^2)\frac{2\epsilon}{a\sqrt{|a|-\epsilon^2}}\qquad (\epsilon>0).
\end{eqnarray}
We remark that integral $I$ is not defined for $x<0$ and $a=-\epsilon^2$ since in this case the integrand blows up as $|x'|^{-3/2}$ near the origin. Moreover, $\Re(x+\ii\epsilon)/\sqrt{a-(x+\ii\epsilon)^2}$ is, for $a<-\epsilon^2$, a discontinuous function of $x$ at $x=0$ because of the cut in the principal determination of the square root. Equation (\ref{eq:epsfinite}) can be verified by taking derivatives with respect to $x$ of both sides: then, the Dirac term created by the latter discontinuity and that coming from the rightmost term cancel out mutually in the rhs. Owing to the delta-sequence
\begin{equation}
\label{eq:dirac1}
\lim_{\epsilon\to 0^+}\frac{\epsilon\,\theta(-a-\epsilon^2)}{a\sqrt{|a|-\epsilon^2}}=-\pi\delta(a),
\end{equation}
one finds $\lim_{\epsilon\to 0}\Re I=-\frac{x}{a}(a-x^2)_+^{-1/2}
-2\pi\theta(-x)\delta(a)$. Using this result in (\ref{eq:fp32}) with $a=\cS^2 t^2-y^2$, shows that $u_i$ in the dislocation formulation (\ref{eq:udisl2}) again yields (P28). Thus, this alternative derivation produces the same result, which illustrates the consistency of the approach.

\subsection{The Dirac term in Equ.\ (P49)}
\label{sec:dirac}
Turning now to (P49),\cite{NOTE1} Markenscoff's objections against the Dirac term $\delta(\xi-x)\dot{\xi}$, where $\xi(t)$ is the dislocation position, misunderstand its nature and working. In the formulation of (P49), mathematically correct in the sense of generalized functions although perhaps inconvenient for numerical calculations, the Dirac term (which comes out of $-\partial\eta/\partial t$) compensates for the singularity as $\tau\to t$ and makes the result finite in the subsonic regime. It serves the same purpose as the subtraction and addition of a compensating term in (M9). A notable difference is that it arises here as a direct consequence of the formalism employed.\cite{PELL10a} Regularizing devices are not unique, and (M9) is by no means the sole manner of writing the elastic strain. In (P49), the current instant $\tau=t$ should be approached in the integral by a limiting process, for instance by multiplying the integrand by $e^{-\epsilon/(t-\tau)}$ and letting $\epsilon\to 0$. This is related to the equal-time value of the Green's function $G_{33}(\mathbf{x}-\mathbf{x'},t-t')$ being defined only as a limit\cite{BART89} $t\to t^{\prime +}$ (instead, considering the out-of-plane stress at small $y\not=0$ would induce a natural time cut-off of the corresponding integral of order $\epsilon \sim |y|/\cS$). With this precaution one can, e.g., deduce from (P49) the stress on the slip plane generated by a Volterra screw that at $t=0$ jumps from rest to a constant velocity $v$, a prototypical instationary case. The result reads\cite{PELL10b}
\begin{eqnarray}
\label{eq:voltjump}
\frac{2\pi}{\mu b}\sigma^{\text{Volt}}(x,t)&=&\frac{1}{x}\biggl[1+\frac{v}{\cS}
\pv\frac{(\cS^2 t^2-x^2)^{1/2}_+}{x-v t}\biggr]\nonumber\\
&-&\pi\mathop{\text{sign}}(v)[(v/\cS)^2-1]^{1/2}_+\delta(x-vt).
\end{eqnarray}
There is \emph{no} Dirac term at the dislocation position in the subsonic regime $|v|<\cS$, as must be. The above-discussed compensation has occurred in the course of the derivation.\cite{PELL10b} Again, the criticism in (M) is unjustified. Had the term $-\partial\eta/\partial t$ not been present in (P35), and consequently no Dirac term featured by (P49), a spurious stress contribution $+\pi (v/\cS)\delta(x-vt)$ would have remained in (\ref{eq:voltjump}). It should be noted that (\ref{eq:voltjump}) embodies a former result for the subsonic regime by Markenscoff\cite{MARK80} [Equ.\ (17) in that reference] which it almost matches, and one for the supersonic regime by Callias and Markenscoff\cite{CALL80} [Equ.\ (2) at $z=0$ in that reference]. I write ``almost'' because the calculation\cite{PELL10b} from (P49) to (\ref{eq:voltjump}) \emph{automatically} provides the principal value prescription $\mathop{\text{p.v.}}$ in the first bracketed term of (\ref{eq:voltjump}), which is absent from Markenscoff's subsonic expression.\cite{MARK80} This prescription regularizes the vicinity of the dislocation position $x=vt$. It makes the elementary stress $\sigma^{\text{Volt}}$ a well-defined generalized function suitable to convolution with a smooth core shape function (remark that $\sigma^{\text{Volt}}$ is regular at $x=0$ for $t>0$), as in statics ($v=0$). Indeed, it is well-known that the Volterra stress kernel in the static PN equation is proportional to $\mathop{\text{p.v.}}1/x$, and not to $1/x$ (or $1/r$) as stated in (M).

\subsection{The matter of Equ.\ (M11)}
This leads us to equation (M11). It originates from the strain of a Volterra screw moving at velocity $v$ being pretended not integrable at $x=vt$.\cite{MARK01} \emph{This cannot be}, owing to the principal value prescription that should be there (see previous Section), and (M11) should be an equality, rather than a non-equality, as physics commands (otherwise, even Weertman's stationary PN equation would be meaningless). To illustrate this point, a straightforward calculation shows that the convolution of $\sigma^{\text{Volt}}$ as given by (\ref{eq:voltjump}) with the function ($\varepsilon>0$) $f_\varepsilon(x)=\pi^{-1}\varepsilon/(x^2+\varepsilon^2)$ is
\begin{eqnarray}
\label{eq:voltjumpa}
\frac{2\pi}{\mu b}[\sigma^{\text{Volt}}*f_\varepsilon](x,t)&=&\Re\biggl\{\frac{1}{x+\ii\varepsilon}\biggl[1\nonumber\\
&&{}+\frac{v}{\cS}
\frac{\sqrt{\cS^2 t^2-(x+\ii\varepsilon)^2}}{(x+\ii\varepsilon)-v t}\biggr]\biggr\}.
\end{eqnarray}
By taking $\varepsilon$ a small number, (\ref{eq:voltjumpa}) provides a numerical representation of (\ref{eq:voltjump}) for any velocity. Moreover, the rhs.\ of (\ref{eq:voltjumpa}), multiplied by $-b/(2\pi)$ (the minus sign being due to a different choice of dislocation sign), coincides with the sum of expression (74) in Ref.\ \onlinecite{MARK01} and of $-[b/(2\pi)]x/(x^2+\varepsilon^2)$ in the subsonic regime for which this expression (74) holds. This sum of terms stands for $\lim_{y\to 0}[\partial_y u_3^{\text{Volt}}(\cdot,y,t)*f_\varepsilon](x)$.\cite{MARK01} This illustrates the fact that in general, and contrary to what (M11) states, the slip-plane limit can be interchanged with the convolution of a Volterra solution by a smooth core function, provided the Volterra solution is interpreted as a generalized function as in (P), in Ref.\ \onlinecite{PELL10b}, and above.

\section{Conclusion}
As far as explicit results are concerned, \emph{all} of the particular consequences worked out so far  from the dynamic PN equations in (P) are in agreement with results obtained by other methods. Comment (M) ignores the generalized-function character of the Volterra solutions that legitimates the approach used. The criticisms against results in (P) have been convincingly refuted by means of simple explicit examples.



\begin{thebibliography}{99}
\bibitem{NOTE1}
In this equation, the square of $\overline{v}_0$ should apply to $\overline{v}_0$ under the square root. Other misprints in  Ref.\ \onlinecite{PELL10a} are as follows: In Sec.\ I, fourth paragraph, `Appendix B' should read `Appendix C1'; also, in Equ.\ (C15), the factor $\cL/\cS^2$ (which actually applies to climb edges) should read $1/\cS$, see (C13).
\end{thebibliography}
\end{document}